\documentclass{aastex631}

\usepackage{amsthm,amsmath,amssymb}
\usepackage{lipsum}
\usepackage{float}
\usepackage{soul}

\begin{document}

\title{The BSN Application-I: Photometric Light Curve Solutions of Contact Binary Systems}

\author[0000-0001-9746-2284]{Ehsan Paki}
\affiliation{Binary Systems of South and North (BSN) Project, Tehran, Iran}

\author[0000-0002-0196-9732]{Atila Poro}
\altaffiliation{atila.poro@obspm.fr}
\affiliation{Binary Systems of South and North (BSN) Project, Tehran, Iran}

\author[0009-0008-3772-5044]{Minoo Dokht Moosavi Rowzati}
\affiliation{Independent Researcher, Tehran, Iran}

\begin{abstract}
Light curve analysis of W UMa-type contact binary systems using MCMC or MC methods can be time-consuming, primarily because the repeated generation of synthetic light curves tends to be relatively slow during the fitting process. Although various approaches have been proposed to address this issue, their implementation is often challenging due to complexity or uncertain performance. In this study, we introduce the BSN application, whose name is taken from the BSN project. The application is designed for analyzing contact binary system light curves, supporting photometric data, and employing an MCMC algorithm for efficient parameter estimation. The BSN application generates synthetic light curves more than 40 times faster than PHOEBE during the MCMC fitting process. The BSN application enhances light curve analysis with an expanded feature set and a more intuitive interface while maintaining compliance with established scientific standards. In addition, we present the first light curve analyses of four contact binary systems based on the TESS data, utilizing the BSN application version 1.0. We also conducted a light curve analysis using the PHOEBE Python code and compared the resulting outputs. Two of the target systems exhibited asymmetries in the maxima of their light curves, which were appropriately modeled by introducing a cold starspot on one of the components. The estimated mass ratios of these total-eclipse systems place them within the category of low mass ratio contact binary stars. The estimation of the absolute parameters for the selected systems was carried out using the $P-a$ empirical relationship. Based on the effective temperatures and masses of the components, three of the target systems were classified as A-subtype, while TIC 434222993 was identified as a W-subtype system.
\end{abstract}


\section{Introduction}
W Ursae Majoris (W UMa) contact binary systems consist of two main-sequence components, typically of F to K spectral types. The components of these systems possess a common convective envelope and have their Roche lobes filled (\citealt{1959cbs..book.....K}). The primary and secondary eclipses exhibit approximately similar minimum depths, indicating that the temperature difference between the stars is slight (\citealt{kuiper1941}, \citealt{2014ApJS..212....4Q}). The stars in contact binary systems transfer mass and energy to one another (\citealt{1979ApJ...231..502L}), leading to potential variations in their orbital periods. These systems are classified into two subtypes: in the A-subtype, the more massive component exhibits a higher effective temperature, whereas in the W-subtype, the less massive component is hotter (\citealt{1970VA.....12..217B}). Determining the mass ratios of some contact systems spectroscopically is challenging due to the broadening and blending of spectral lines. Photometric analysis is an effective method for studying systems in which one of the stars has a small mass and radius, as estimating the mass ratio from spectroscopic data may be challenging (\citealt{2022AJ....164..202L}, \citealt{2024AA...692L...4L}, \citealt{2025MNRAS.538.1427P}). However, in systems with total eclipses, photometric analysis yields more accurate mass ratios than in systems with partial eclipses (\citealt{2021AJ....162...13L}, \citealt{2024AJ....168..272P}).

In recent decades, different programs and codes have been developed to analyze the light curves of binary systems, and a variety of methods have been applied to estimate the fundamental parameters of the stars. Currently, most light curve analyses are conducted using the Wilson-Devinney (\citealt{1971ApJ...166..605W}) and the PHysics Of Eclipsing BinariEs (PHOEBE, \citealt{2016ApJS..227...29P}) programs, which offer the capability to implement Monte Carlo (MC) simulations and the Markov Chain Monte Carlo (MCMC) method, respectively. MC simulations and the MCMC method are valuable tools for light curve analysis, providing reliable uncertainty estimates. However, the application of these methods can be time-consuming. On regular computers, the search process can take several days. Furthermore, if there is a mistake in the distribution determination or if the final result is unsatisfactory, the entire process needs to be repeated. As a result, light curve analysis can be time-consuming, and with the growing volume of data from space telescopes, there is a need to propose efficient solutions.

Recent studies, such as \cite{2023MNRAS.525.4596D} and \cite{2025ApJS..277...51L} have attempted to accelerate the MCMC process in the PHOEBE code by incorporating neural network (NN) models. These studies achieved significant improvements in the speed of the MCMC process, enabling the automated generation of light curve solutions for a large number of contact binary systems. The use of NNs to generate models that significantly accelerate the light curve solution process has also been explored in other studies (e.g., \citealt{2025AA...693A.131W}). Although the use of methods such as neural networks (NNs) is highly valuable, these approaches still encounter significant challenges, as discussed in the study by \cite{2025MNRAS.537.3160P}.

We developed the BSN application under the Binary Systems of South and North (BSN) project. The application was evaluated using over 80 previously studied contact binary systems with available data, and the results were compared with those obtained using PHOEBE. To further demonstrate the capabilities of the BSN application and assess its performance, we also applied it to perform the first light curve analyses of four contact binary systems. The general characteristics of target systems are listed in Table \ref{info} from the Gaia DR3 ({\url{https://gea.esac.esa.int/archive/}}), the All-Sky Automated Survey for Supernovae (ASAS-SN; \citealt{2018MNRAS.477.3145J}), and TESS Input Catalog (TIC; \citealt{2018AJ....156..102S}) v8.2 databases. The target contact binary systems have orbital periods ranging from 0.4 to 0.5 days, with apparent magnitudes between $13.15^{mag.}$ and $13.81^{mag.}$ based on the ASAS-SN catalog. In this study, we used the Transiting Exoplanet Survey Satellite (TESS) data for light curve analysis. The data are publicly available from the Mikulski Archive for Space Telescopes (MAST ({\url{http://archive.stsci.edu/tess/all\_products.html}})). The TESS sector and exposure length used for each system are listed in Table \ref{info}.

\begin{table*}
\centering
\renewcommand\arraystretch{1.5}
\caption{Characteristics of target systems.}
\begin{tabular}{c|ccc|cc|cc}
\hline
\textbf{System} & \multicolumn{3}{c|}{\textbf{Gaia DR3}} & \multicolumn{2}{c|}{\textbf{ASAS-SN}} & \multicolumn{2}{c}{\textbf{TESS}}\\
& \textbf{RA.(J2000)} & \textbf{DEC.(J2000)} & \boldmath$d$\textbf{(pc)} & \boldmath$V$\textbf{(mag.)} & \textbf{P(day) }& \textbf{Sector} & \textbf{Exposure Length(s)}\\
\hline
TIC 406453269	& 47.112188 & $-$16.151984 & 799 (9) &	13.15 &	0.4763368 &	31 & 600\\
TIC 423750361	& 29.797049 & $-$17.728352 & 1461 (39) & 13.81	& 0.4477891 &	30 & 600\\
TIC 434222993	& 10.956649 & 18.054733 & 1081 (20) & 13.73 & 0.4625365 &	57 & 200\\
TIC 71743300	& 25.856993 & 55.722065 & 1028 (19) & 13.76 & 0.4396651 &	58 & 200\\
\hline
\end{tabular}
\label{info}
\end{table*}

\vspace{6pt}
\section{Light Curve Solutions}
\subsection{BSN Application}
The BSN application is a desktop tool developed using the latest .NET platform. It features a graphical user interface (GUI) designed to streamline user interaction throughout the modeling process. Scientific-based and performance details for this application are provided as follows.

$\textbf{(A) Phase Calculation Methodology:}$ Although the BSN application performs all calculations in time, representing results in orbital phase is often more practical for interpreting eclipsing binary behavior. Phases are mapped to the ($-0.5$, $0.5$) interval, and changes in orbital period over time are accounted for in the time-to-phase conversion. A phase shift parameter helps align synthetic and observed curves but does not affect the phasing of observations.

$\textbf{(B) Three-dimensional modeling:}$ \cite{1968ApJ...153..877L} projected the stellar surface onto the sky plane, treating the depth along the line of sight as an unknown to be resolved separately. Subsequently, \cite{1971ApJ...166..605W} implemented a spherical coordinate system, while \cite{1972MNRAS.156...51M} opted for cylindrical coordinates to describe the geometry. PHOEBE incorporates a simplified spot model based on the \cite{1971ApJ...166..605W} method.

A three-dimensional image of the system is displayed on the main screen of the BSN application, with the option to adjust the relative positions of the stars during different phases. The geometric shapes of the stars in the BSN application were computed using the method proposed by \cite{1984ApJS...55..551M}.

$\textbf{(C) Effective surface temperature:}$ In the BSN application, local effective temperatures across a star's surface are primarily influenced by tidal and rotational distortions and are modeled using a power-law gravity-darkening approach. The local temperature in each region is derived from the polar effective temperature, which is itself calculated from the global mean effective temperature, weighted by surface area and local gravity. In contact binaries, the stellar surfaces are divided using a boundary plane at the neck, with surface areas assigned to each component based on their location relative to this boundary. Temperature discontinuities can arise at the neck if the components have different polar temperatures, making this method best suited for systems in or near thermal equilibrium. Additionally, the BSN application includes a basic spot model, where localized surface temperature variations are introduced via circular regions defined by angular coordinates and a temperature scaling factor.

$\textbf{(D) Bolometric albedo and gravity darkening:}$ The bolometric albedo and gravity-darkening coefficients are considered in this application. The user can apply the values assumed in the application based on the studies \cite{1969AcA....19..245R} and \cite{1967ZA.....65...89L} or enter other values.

$\textbf{(E) Limb darkening:}$ Limb-darkening coefficients play a crucial role in the modeling of contact binary stars, where both components share a common envelope and exhibit significant geometric distortion due to mutual gravitational interaction (\citealt{1992AA...259..227D,1993AJ....106.2096V}). The limb-darkening effect, which causes the stellar intensity to decrease from the center to the edge of the disk, is typically described using analytical laws—such as linear, quadratic, square-root, or logarithmic formulations—with coefficients dependent on local effective temperature, surface gravity, and wavelength. In the BSN application version 1.0, we employed linear and logarithmic limb-darkening formulations, using coefficients taken from the tabulations provided by \cite{1993AJ....106.2096V}. Accurate treatment of limb darkening is essential, as it directly influences the determination of key system parameters, including inclination, temperature ratio, and component radii, thereby affecting the reliability of the derived physical and evolutionary characteristics of the binary system.

It is worth noting that the method introduced by \cite{1990ApJ...356..613W} for modeling the reflection effect in binary star systems required effective wavelength-dependent limb-darkening coefficients, as well as bolometric coefficients, which were incorporated into bolometric analogues of the relevant equations. Accordingly, the bolometric coefficients calculated by \cite{1993AJ....106.2096V} are used for modeling the reflection effect in binary systems within the BSN application.

$\textbf{(F) Stellar atmosphere modeling:}$ Local emergent intensity depends on a range of intrinsic stellar properties such as temperature, surface gravity, chemical composition, and rotation. While a simplified approach treats stars as blackbodies based solely on temperature, accurate modeling requires assigning appropriate model atmospheres based on stellar type and application.

The BSN application models the stellar atmosphere using the method of \cite{2004AA...419..725C}. The most comprehensive atmospheric model currently available and the one used in PHOEBE is presented by \cite{2004AA...419..725C}. Since the BSN application employs the method of \cite{2004AA...419..725C} for modeling stellar atmospheres, the lower temperature limit of 3500 K is imposed by this method. The upper limit in the BSN application is 8500 K, which is suitable for the temperature range of most contact binary systems.

$\textbf{(G) Passband:}$ The BSN application is capable of performing light curve analysis using the standard Johnson $UBVR_cI_c$ filters (\citealt{1965ApJ...141..923J}) and the TESS passband. Therefore, the application allows users to model each dataset separately, using its corresponding filter.

In stellar photometric modeling, the emergent intensity within a specific passband is derived by integrating the Spectral Energy Distribution (SED), perpendicular to the stellar surface, weighted by the passband transmission function. Specifically, this intensity is obtained by multiplying the SED by the transmission function and integrating over wavelength. Direct integration for every combination of stellar parameters-such as effective temperature, surface gravity, and metallicity-is computationally intensive. To mitigate this, precomputed intensity values based on a broad grid of model atmospheres (\citealt{2004AA...419..725C}) are stored in advance. Interpolation over this grid in \(T_{\text{eff}}\), \(\log g\), and [M/H] allows for efficient and accurate estimates. Although microturbulent velocity (\(v_{\text{turb}}\)) generally influences spectral line formation, its effect on passband-integrated intensities is negligible and can be omitted. This is because increasing \(v_{\text{turb}}\) broadens and shallows spectral lines without significantly altering their equivalent widths, thereby leaving the integral essentially unchanged. Any residual effects near the edges of the passband are minimal due to the gradual decline in transmission efficiency.

Since observational data are rarely provided in absolute intensity units, passband luminosity is used to rescale model intensities for comparison with observations. This involves integrating the emergent intensities across the stellar surface using a limb-darkening model to convert from intensity to flux, yielding the total luminosity within the passband. This computed luminosity is then used to normalize the intensities.

$\textbf{(H) Reflection effect:}$ Among the various challenges in computing theoretical light curves for contact binary stars is the need to account for mutual irradiation between the components, commonly referred to as the reflection effect. The BSN application incorporates the reflection effect into light curve modeling using the method proposed by \cite{1990ApJ...356..613W}. However, because this effect is minimal in most contact binary systems, it may be disregarded to improve computational performance without significantly impacting accuracy.

$\textbf{(J) MCMC:}$ The BSN application offers a high-speed MCMC process that operates efficiently even on personal computers with standard CPUs. The MCMC process in the BSN application operates on five main parameters: $i$, $q$, $f$, $T_1$, and $T_2$.

We employed the affine-invariant ensemble sampler, known as the single stretch move algorithm, for MCMC, which underlies the widely used \texttt{emcee} package and was originally introduced by \cite{2010CAMCS...5...65G}. This algorithm significantly outperforms standard Metropolis–Hastings (M–H) methods by producing independent samples with much shorter autocorrelation times (\citealt{2013PASP..125..306F}).

\vspace{0.6cm}
\subsection{Light Curve Solution of Four Target Systems}
The first light curve modeling of four target systems was conducted in contact mode, based on the TESS-observed light curve shapes, catalog classifications, and the systems' short orbital periods. The gravity-darkening coefficient ($g_1=g_2=0.32$), the bolometric albedo ($A_1=A_2=0.5$), the stellar atmosphere model, and limb-darkening coefficients were applied according to the assumptions of the BSN application.

The initial effective temperature ($T$) used in the analysis was taken from the Gaia DR3 database. Based on the depth of the minima in the light curves, the hotter component was identified, and the Gaia DR3 temperature was adopted for this star. The initial effective temperature of the cooler star was estimated from the depth difference between the primary and secondary minima in the light curves.

The analysis indicates that TIC 406453269 and TIC 71743300 exhibit noticeable asymmetries in the maxima of their light curves. This phenomenon, commonly known as the O'Connell effect (\citealt{1951PRCO....2...85O, 2017AJ....153..231S}), is typically attributed to magnetic activity on the stellar surfaces that leads to the formation of starspots. To account for these asymmetries, a cold starspot was introduced on one of the components. Although the initial modeling considered a starspot on the secondary component, placing it on the primary component resulted in a significantly improved fit. It is worth noting that the secondary stars have significantly smaller radii. Since such asymmetries may also result from transient phenomena such as flares, the full span of the TESS sector data was examined; however, no evidence of such events was found. Furthermore, in the absence of spectroscopic data and indicators of chromospheric activity, the starspot model was employed purely to enhance the photometric fit. The parameters used for the starspot model—including colatitude (Col.$^\circ$), longitude (Long.$^\circ$), angular radius (Radius$^\circ$), and temperature ratio ($T_{\mathrm{spot}}/T_{\mathrm{star}}$)—are provided in Table \ref{analysis}.

\begin{table*}
\centering
\renewcommand\arraystretch{1.8}
\caption{Light curve solution results.}
\begin{tabular}{C C C C C}
\hline
\textbf{Parameter} & \textbf{TIC 406453269} & \textbf{TIC 423750361 }& \textbf{TIC 434222993} & \textbf{TIC 71743300}\\
\hline
$T_{1}$ (K) & $5650_{\rm-(40)}^{+(41)}$ & $6722_{\rm-(45)}^{+(46)}$ & $5945_{\rm-(43)}^{+(43)}$ & $6536_{\rm-(51)}^{+(45)}$\\

$T_{2}$ (K) & $5607_{\rm-(43)}^{+(41)}$ & $6419_{\rm-(43)}^{+(48)}$ & $5715_{\rm-(42)}^{+(43)}$ & $6514_{\rm-(44)}^{+(45)}$\\

$q=M_2/M_1$ & $0.173_{\rm-(7)}^{+(8)}$ & $0.204_{\rm-(10)}^{+(12)}$ & $4.799_{\rm-(96)}^{+(99)}$ & $0.197_{\rm-(15)}^{+(15)}$\\

$i^{\circ}$	& $82.27_{\rm-(2.03)}^{+(1.55)}$ & $85.86_{\rm-(1.91)}^{+(1.81)}$ & $84.04_{\rm-(1.74)}^{+(1.68)}$ & $80.51_{\rm-(1.70)}^{+(1.66)}$\\

$f$ & $0.586_{\rm-(67)}^{+(77)}$ & $0.317_{\rm-(79)}^{+(79)}$ & $0.595_{\rm-(69)}^{+(69)}$ & $0.595_{\rm-(86)}^{+(90)}$\\

$\Omega_1=\Omega_2$ & 2.099(56) & 2.201(65) & 8.536(318) & 2.150(63)\\

$l_1/l_{tot}$ & 0.821(11) & 0.824(13) & 0.232(7) & 0.801(13)\\

$l_2/l_{tot}$ & 0.179(6) & 0.176(6) & 0.768(12) & 0.199(6)\\

$r_{(mean)1}$ & 0.560(18) & 0.536(14) & 0.285(26) & 0.551(20)\\

$r_{(mean)2}$ & 0.269(23) & 0.268(16) & 0.547(21) & 0.280(24)\\
\hline										
$Col.^\circ$ (spot) & - & 86 & - & 86\\
$Long.^\circ$ (spot) & - & 57 & - & 302\\
$Radius^\circ$ (spot) & - & 14 & - & 16\\
$T_{spot}/T_{star}$ & - & 0.94 & - & 0.90\\
Component & - & Primary & - & Primary\\
\hline
\end{tabular}
\label{analysis}
\end{table*}

Photometric data, along with initial parameter values, were utilized to achieve a satisfactory theoretical fit. The optimization tool within the BSN application was also employed to improve the light curve solutions. The optimization process was carried out for five primary parameters ($T_1$, $T_2$, $q$, $f$, $i$) and four starspot characteristics.

\begin{figure*}
\centering
\includegraphics[scale=0.4]{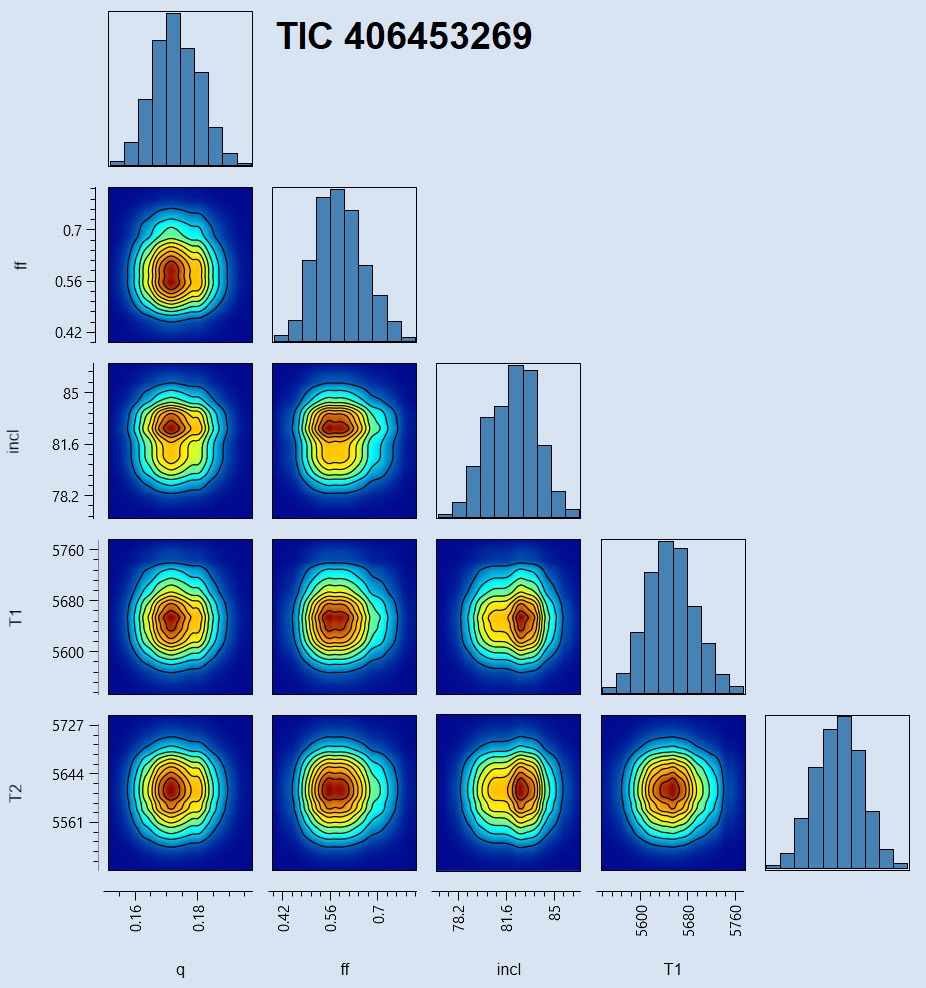}
\includegraphics[scale=0.4]{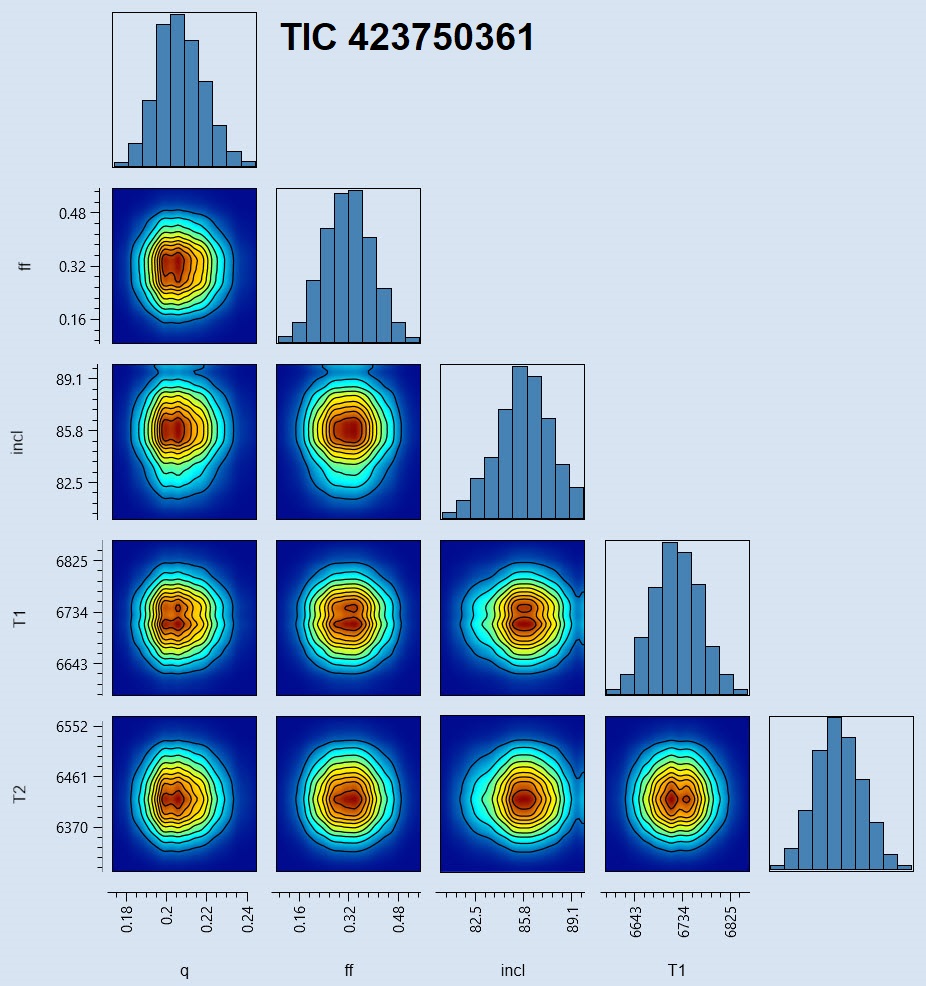}
\includegraphics[scale=0.4]{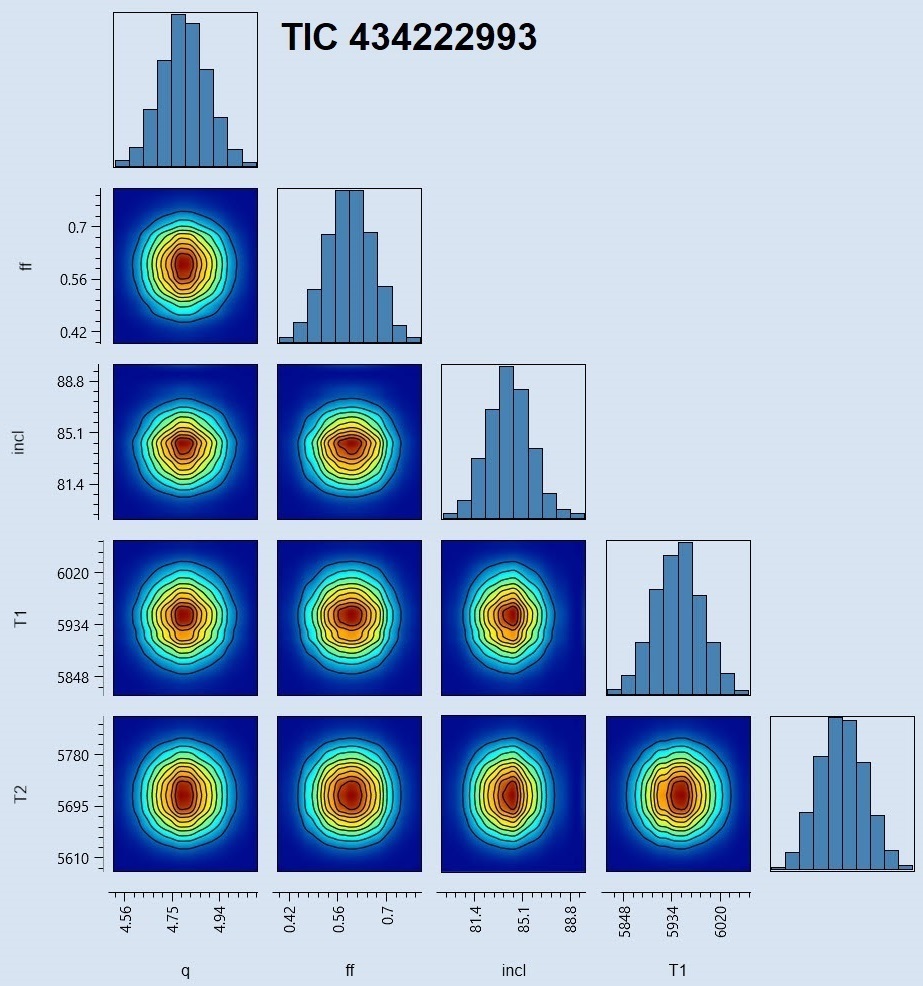} 
\includegraphics[scale=0.4]{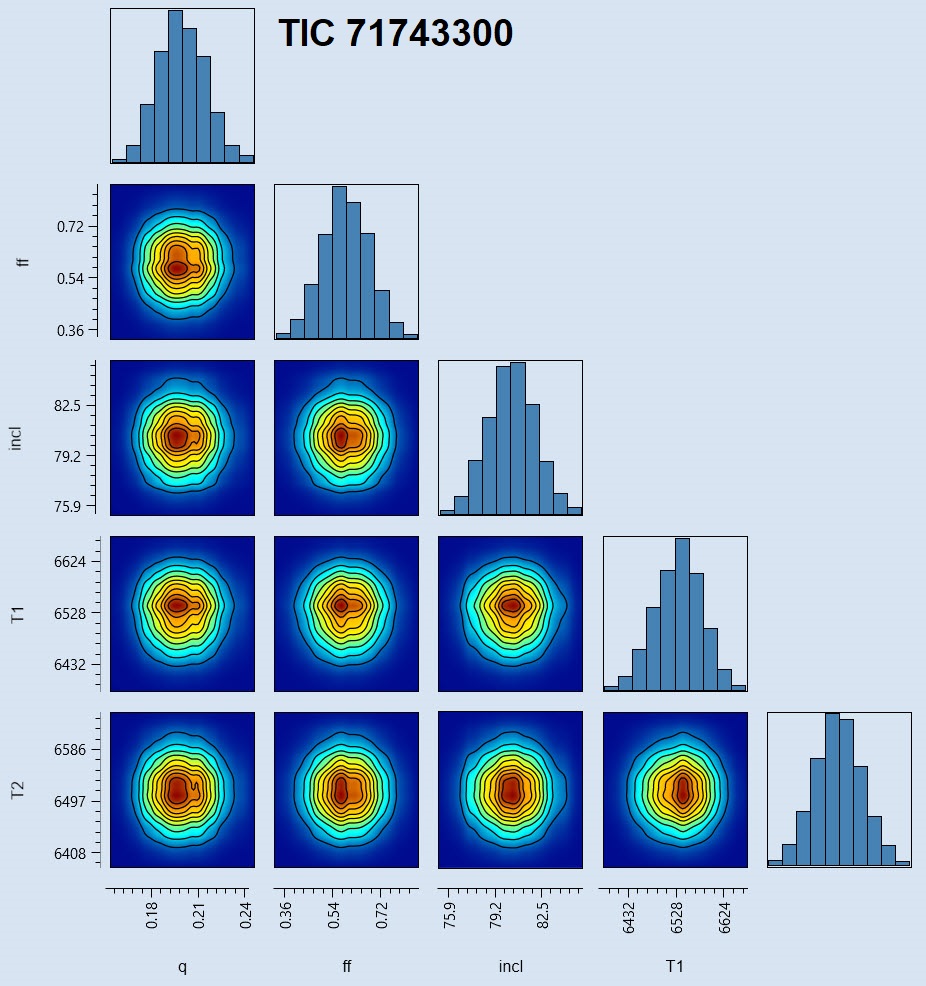}
\caption{Corner plots based on the heat-map of the target contact binary systems were determined by MCMC modeling.}
\label{corner}
\end{figure*}

\begin{figure*}
\centering
\includegraphics[scale=0.3]{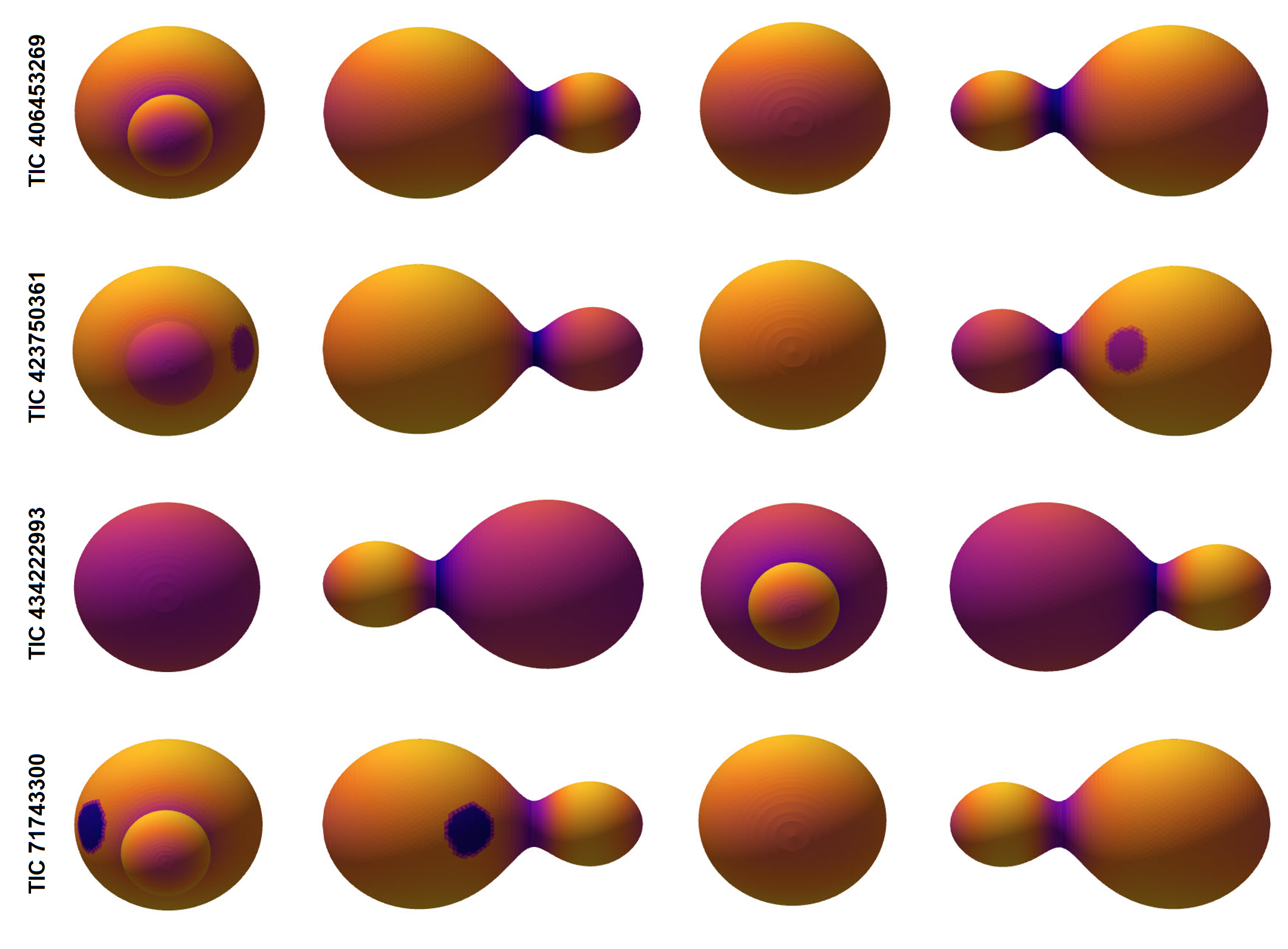}
\caption{3D view of the systems in 0, 0.25, 0.5, and 0.75 phases, respectively.}
\label{3D}
\end{figure*}

The MCMC method was then applied to refine the five estimated parameters ($i$, $q$, $f$, $T_1$, and $T_2$) using the TESS data for each system, and the corresponding uncertainty values were determined. We employed 24 walkers and 1000 iterations for each walker for target contact binary systems. Table \ref{analysis} presents the results of the light curve analysis. For each system, the BSN application generated a corner plot based on the heat-map produced by the MCMC process (Figure \ref{corner}). Figure \ref{3D} shows three-dimensional representations of the binary systems.

It is worth noting that the BSN application is capable of performing model fitting with a third light component ($l_3$). The light curves of the target systems were appropriately modeled without including $l_3$; there is no indication of contaminating light from a nearby star, and no additional evidence currently supports $l_3$ presence. Nevertheless, these factors alone do not definitively rule out the existence of a third light contribution. A more conclusive evaluation would require a combined analysis with the O-C (Observed–Calculated) diagram, which lies beyond the scope of this work.

\begin{figure*}
\centering
\includegraphics[scale=0.23]{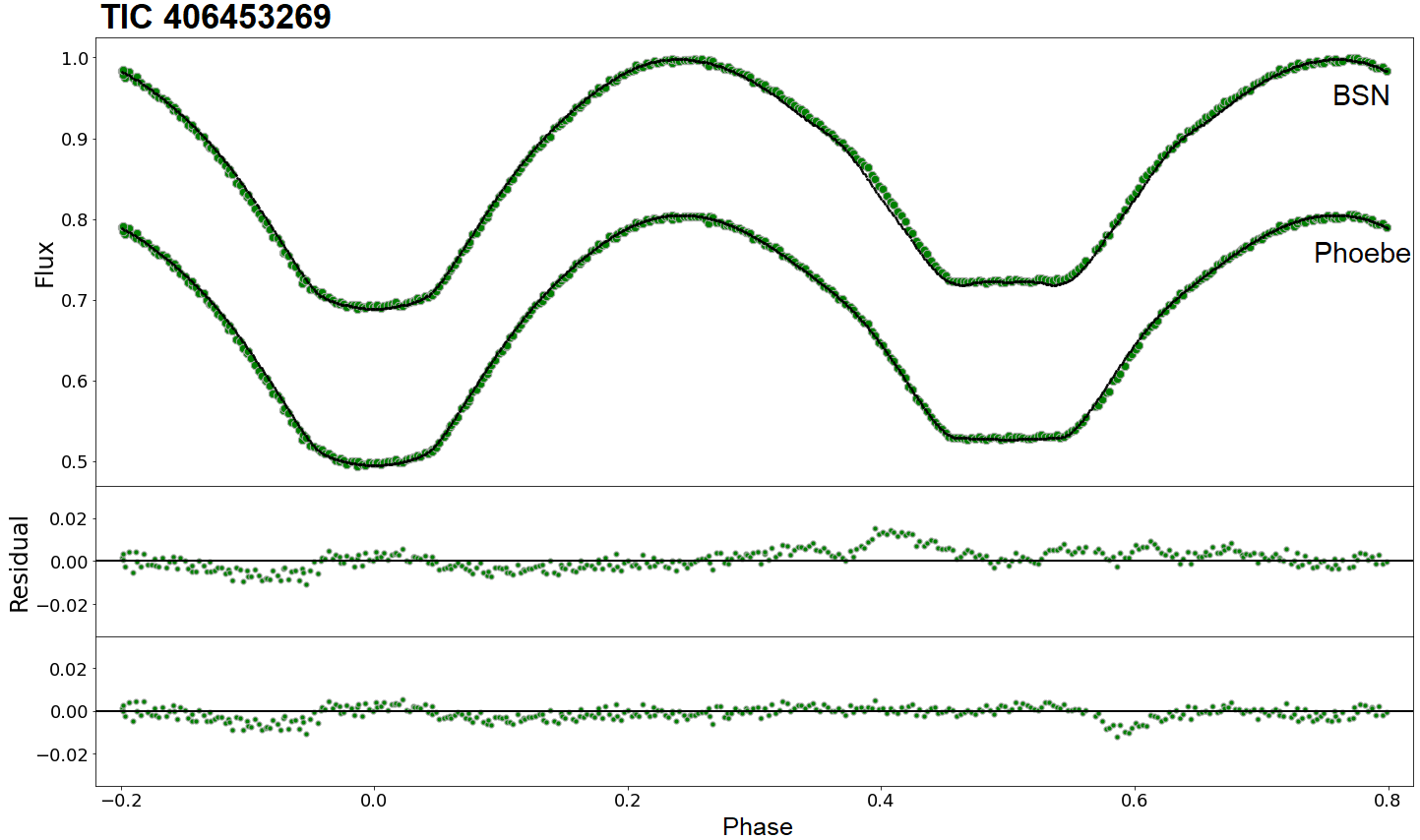}
\includegraphics[scale=0.23]{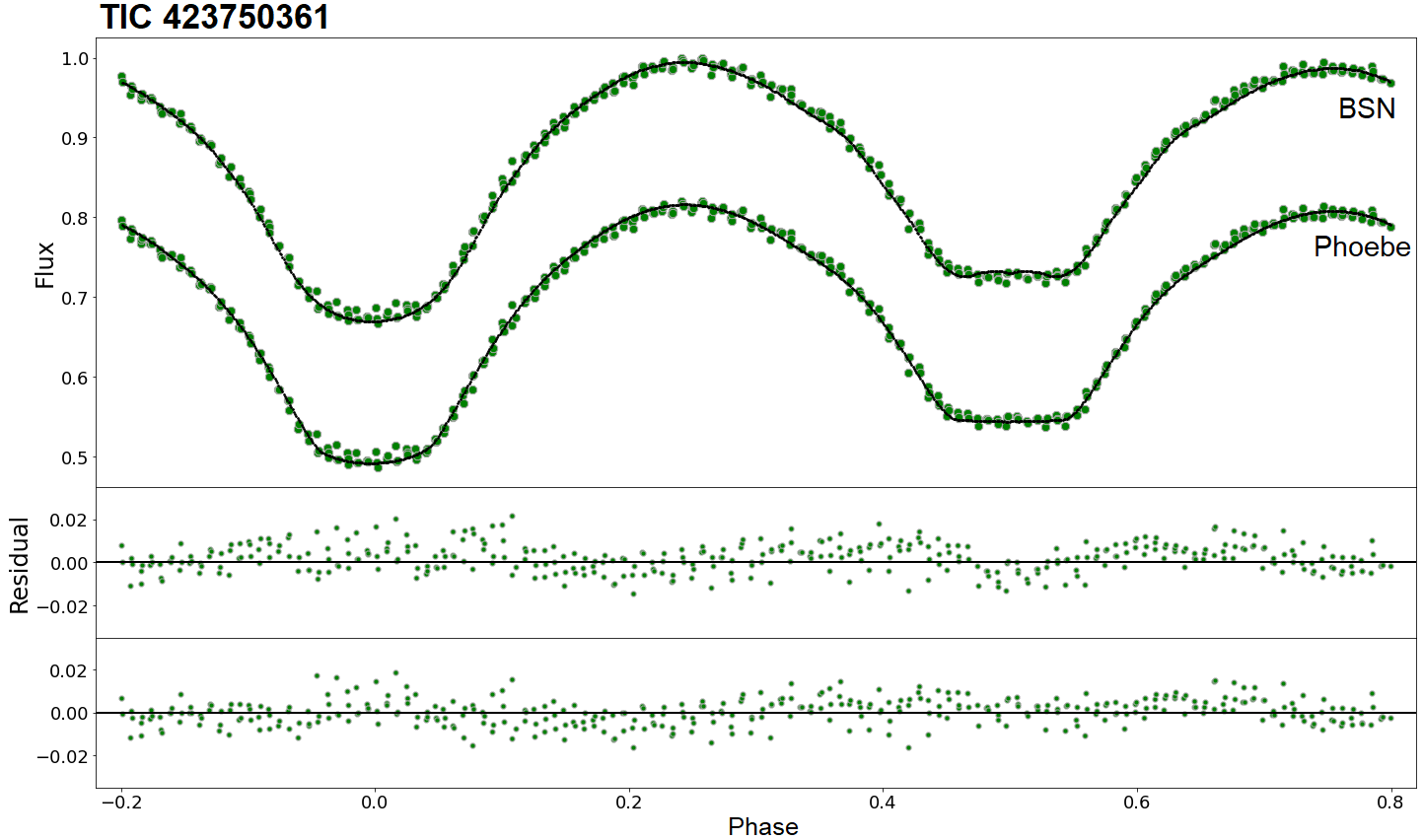}
\includegraphics[scale=0.23]{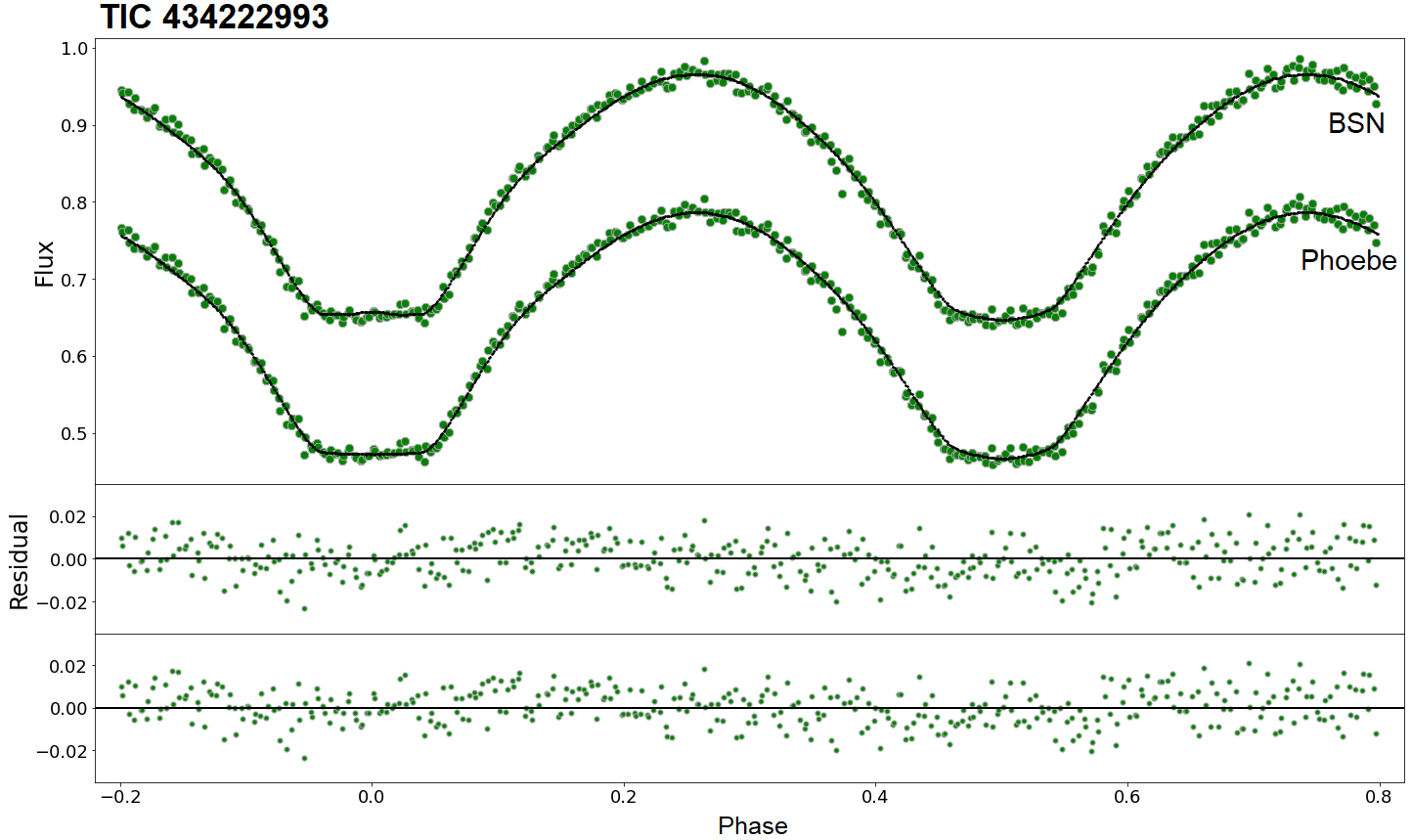}
\includegraphics[scale=0.23]{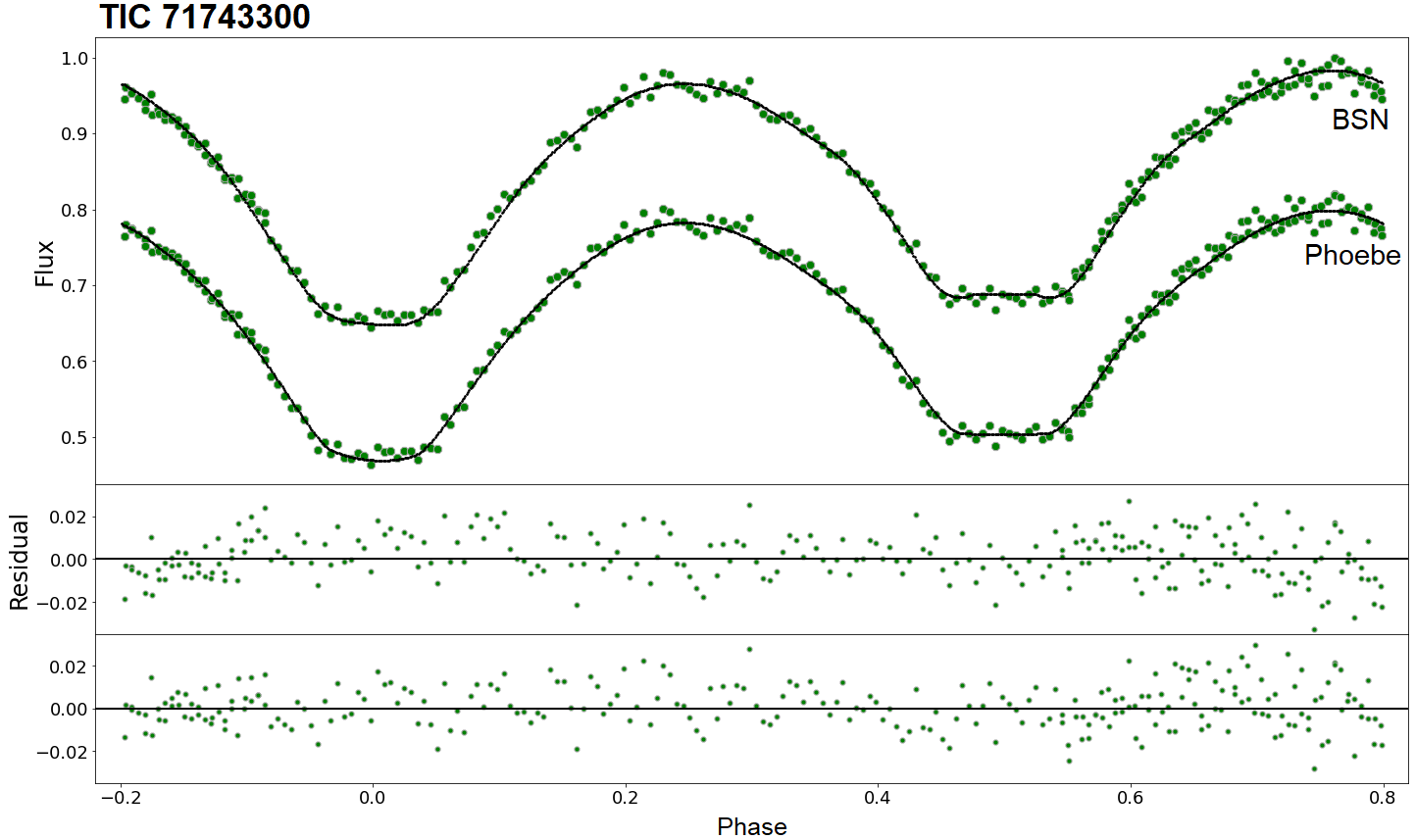}
\caption{TESS observations (green dots) and modeled light curves (black lines) for the target binary systems. The upper fit corresponds to the BSN application, while the lower fit was generated using the PHOEBE program. For better visibility, the flux values of the PHOEBE model light curves have been vertically shifted.}
\label{lc}
\end{figure*}

We examined the final results from the MCMC process of the BSN application using the Python implementation of PHOEBE version 2.4.9. Figure \ref{lc} presents the TESS observational data alongside the synthetic light curves generated by the BSN application and PHOEBE for the target binary systems. As shown in Figure \ref{lc}, the synthetic light curves are nearly identical.

\vspace{6pt}
\section{Estimation of Absolute Parameters}
Several methods are available for estimating the absolute parameters of contact binary systems (\citealt{2024NewA..11002227P,2024RAA....24a5002P}). Since the study was based on photometric data from the TESS band, we employed an empirical relationship between the orbital period and semi-major axis ($P-a$) to estimate the absolute parameters.

The study by \cite{2024PASP..136b4201P} updated the $P-a$ empirical relationship using data from 414 contact binary systems between the orbital periods of 0.2 and 0.7 days (Equation \ref{eqPa}).

\begin{equation}
\label{eqPa}
a=(0.372_{\rm-0.114}^{+0.113})+(5.914_{\rm-0.298}^{+0.272})\times P
\end{equation}

We started the estimation of absolute parameters for the target systems by applying Equation \ref{eqPa}, using the orbital period of each system (Table \ref{info}) to calculate the semi-major axis $a(R_{\odot})$. Then, we estimated the mass ($M$) of each component using the mass ratio derived from the light curve solutions in combination with Kepler's third law, as expressed in Equations \ref{eq:M1} and \ref{eq:M2}.

\begin{eqnarray}
M{_1}=\frac{4\pi^2a^3}{GP^2(1+q)}\label{eq:M1}\\
M{_2}=q\times{M{_1}}\label{eq:M2}
\end{eqnarray}

The radius ($R$) of each component was determined using the mean fractional radii ($r_{mean1,2}$) from Table \ref{analysis} and the $R=a \times r$ relation. The luminosity of each star was calculated using its effective temperature and radius. The absolute bolometric magnitude ($M_{\text{bol}}$) was determined using the logarithmic relationship between luminosity and bolometric magnitude, as expressed in Equation~\ref{eq:Mbol}.

\begin{equation}\label{eq:Mbol}
M_{bol1,2)}=M_{bol\odot}-2.5\times log\left(\frac{L_{1,2}}{L_\odot}\right)
\end{equation}

In Equation \ref{eq:Mbol}, the absolute bolometric magnitude of the Sun is assumed to be $4.73^{\text{mag.}}$, based on the \cite{2010AJ....140.1158T} study. Using the calculated mass and radius values, the surface gravity ($g$) of the stars was estimated on a logarithmic scale. The orbital angular momentum ($J_0$) was also calculated, using Equation \ref{eqJ0} from \cite{2006MNRAS.373.1483E},

\begin{equation}\label{eqJ0}
J_0=\frac{q}{(1+q)^2} \sqrt[3] {\frac{G^2}{2\pi}M^5P}
\end{equation}

\noindent where $q$ is the mass ratio, $M$ is the total mass of the system, $P$ is the orbital period, and $G$ is the gravitational constant. The resulting absolute parameters for the four contact binary systems are provided in Table \ref{absolute}.

\begin{table*}
\centering
\renewcommand\arraystretch{1.5}
\caption{Estimated absolute parameters.}
\begin{tabular}{C C C C C}
\hline
\textbf{Parameter} & \textbf{TIC 406453269} & \textbf{TIC 423750361 }& \textbf{TIC 434222993 }& \textbf{TIC 71743300}\\
\hline
$M_1(M_\odot)$ &	1.636(414)	&	1.532(397)	&	0.325(83)	&	1.523(397)	\\
$M_2(M_\odot)$ &	0.283(88)	&	0.313(102)	&	1.558(439)	&	0.300(107)	\\
$R_1(R_\odot)$ &	1.786(201)	&	1.619(175)	&	0.886(157)	&	1.638(196)	\\
$R_2(R_\odot)$ &	0.858(146)	&	0.809(117)	&	1.700(205)	&	0.832(144)	\\
$L_1(L_\odot)$ &	2.930(805)	&	4.823(1.262)	&	0.883(377)	&	4.412(1.282)	\\
$L_2(L_\odot)$ &	0.656(270)	&	1.003(348)	&	2.778(815)	&	1.124(466)	\\
$M_{bol1}(mag.)$ &	3.563(264)	&	3.022(252)	&	4.865(386)	&	3.118(277)	\\
$M_{bol2}(mag.)$ &	5.188(374)	&	4.727(324)	&	3.621(279)	&	4.603(376)	\\
$log(g)_1(cgs)$ &	4.148(5)	&	4.205(11)	&	4.055(43)	&	4.192(3)	\\
$log(g)_2(cgs)$ &	4.023(19)	&	4.116(6)	&	4.170(9)	&	4.075(6)	\\
$a(R_\odot)$ &	3.189(249)	&	3.020(241)	&	3.107(245)	&	2.972(239)	\\
$logJ_0$ &	51.557(182)	&	51.569(188)	&	51.594(171)	&	51.548(198)	\\
\hline
\end{tabular}
\label{absolute}
\end{table*}

\vspace{6pt}
\section{Discussion and Conclusions}
(A) In this work, we introduced the BSN application, designed for modeling the light curves of contact binary stars, and addressed issues related to the speed of the MCMC process. By offering a more user-friendly and practical interface—such as simplified identification of star spot coordinates when needed—the application allows users to focus more effectively on scientific analysis. The application is currently available for the Windows operating system. Its improved MCMC speed is attributed to the engineering structure and integration of updated tools, while the underlying scientific procedures for light curve analysis remain consistent with those used in other programs. In fact, it was the substantial increase in the speed of synthetic light curve generation that led to the improved performance of the MCMC process in the application.

(B) We conducted the first light curve analysis of four contact binary systems using the BSN application. These solutions were derived based on TESS observations. Among the target systems, TIC 423750361 exhibited the highest temperature difference between its two stars at 303 K, while TIC 71743300 showed the lowest, at 22 K. Additionally, the solutions for the TIC 423750361 and TIC 71743300 systems required the inclusion of starspots to account for the asymmetry observed in the maxima of their light curves. To obtain the best-fit solution, the optimization and MCMC modules of the BSN application were utilized.

The spectral types of the stars were determined using the temperature criteria outlined by \cite{2000asqu.book.....C} and \cite{2018MNRAS.479.5491E}. Accordingly, spectral types for $T_1$ and $T_2$ (Table \ref{analysis}) G7-G7, F2-F5, G1-G5, and F3-F3 were attributed to the systems TIC 406453269, TIC 423750361, TIC 434222993, and TIC 71743300, respectively.

We estimated the absolute parameters using the empirical $P-a$ relationship. Based on the light curve analysis and the estimated absolute parameters of the target systems, three were classified as A-subtype, while TIC 434222993 was identified as W-subtype. This conclusion is based on the classification of contact binaries introduced by \cite{1970VA.....12..217B}. In the A-subtype, the star with greater mass has a higher effective temperature. Conversely, in the W-subtype, the component with lower mass is the hotter one (Tables \ref{analysis} and \ref{absolute}).

The fill-out factors of contact binary systems are categorized into three types: deep ($f\geq 50\%$), medium ($25\% \leq f < 50\%$), and shallow ($f<25\%$) eclipsing contact binaries (\citealt{2022AJ....164..202L}). With the exception of TIC 423750361, which falls into the medium category, all other systems exhibit a deep contact degree.

\begin{figure*}
\centering
\includegraphics[scale=0.35]{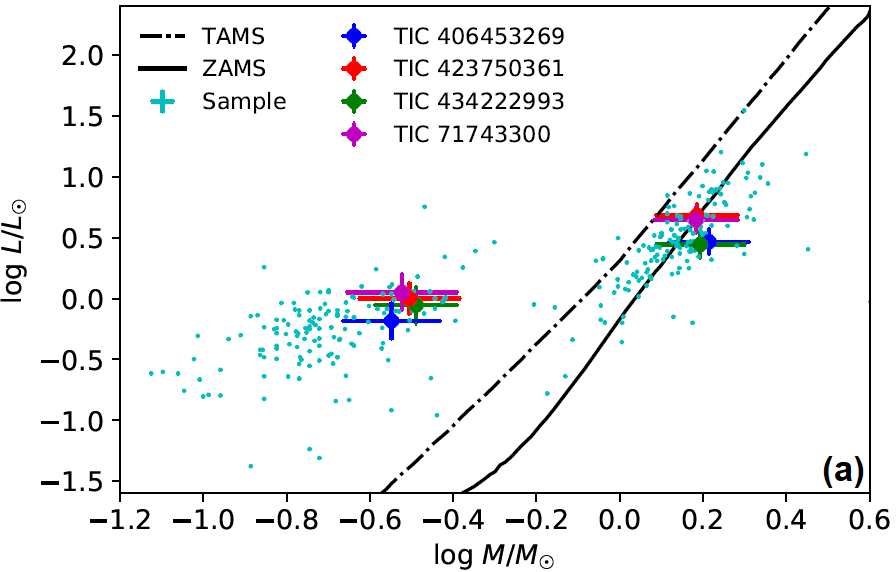}
\includegraphics[scale=0.35]{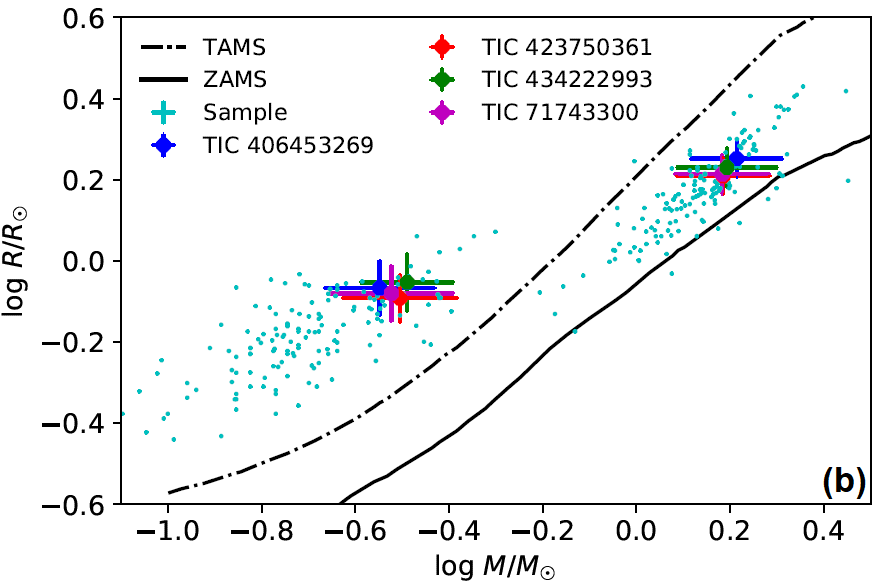}
\includegraphics[scale=0.35]{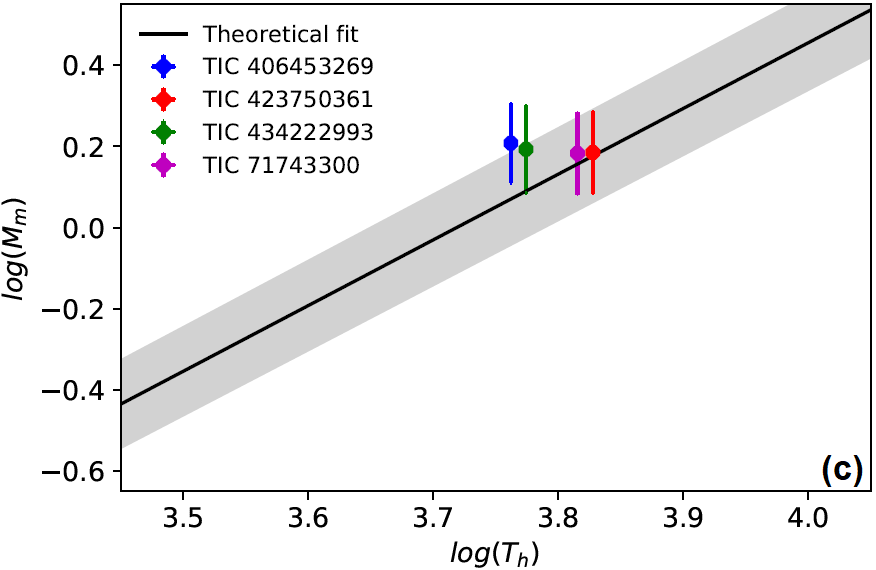}
\includegraphics[scale=0.35]{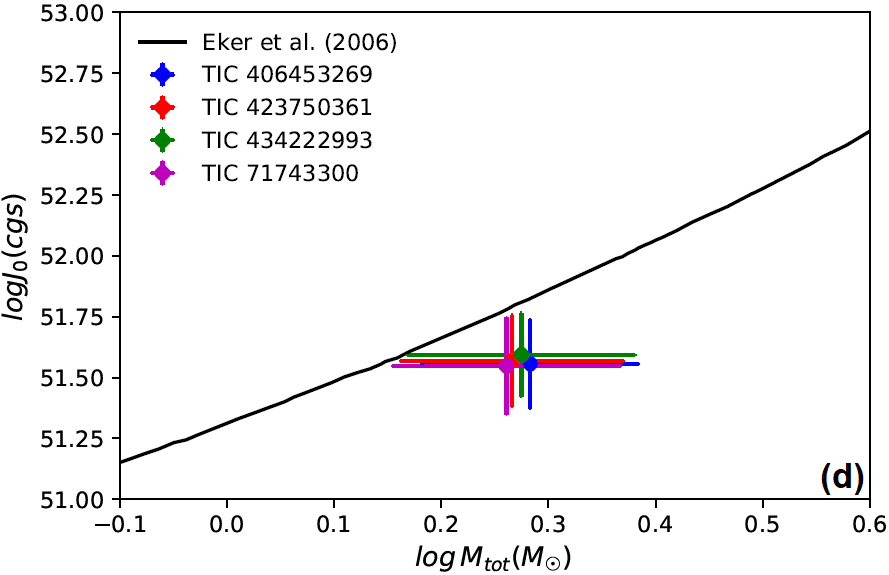}
\caption{(a) $M_{1,2}-L_{1,2}$, (b) $M_{1,2}-R_{1,2}$, (c) $T_h-M_m$, and (d) $M_{tot}-J_0$ diagrams. A sample of 155 low-mass ratio contact binary systems, selected from \cite{2025MNRAS.538.1427P}, is displayed in diagrams (a,b).}
\label{4diagrams}
\end{figure*}

(C) The positions of the stars are displayed in Figure \ref{4diagrams}, with the Zero-Age Main Sequence (ZAMS) and the Terminal-Age Main Sequence (TAMS) lines from the \cite{2000AAS..141..371G} study, on the Mass--Luminosity ($M-L$) and Mass--Radius ($M-R$) diagrams. Based on the light curve solutions, the estimated absolute parameters, and the positions of the components in the $M-L$ and $M-R$ diagrams, the more massive and hotter stars in the three target systems are located near the ZAMS, while the less massive and cooler components lie above the TAMS. In contrast, for the TIC 434222993 system, the hotter and less massive component is positioned above the TAMS line, indicating that this binary follows a distinct evolutionary path compared to the other three systems. Additionally, we selected a sample of 155 binaries from the total of 818 contact systems presented in the \cite{2025MNRAS.538.1427P} study. The selected systems satisfy three criteria: (1) a mass ratio less than 0.25, (2) an orbital period shorter than 0.7 days, and (3) the availability of absolute parameters, including mass, radius, and luminosity. This sample was used as a reference population to compare the target systems of our study in the $M-L$ and $M-R$ diagrams, following the approach of ZAMS and TAMS evolutionary tracks.

The position of the stars above the TAMS line can be explained by mass transfer processes: initially, the more massive component evolves faster and expands to fill its Roche lobe, initiating mass transfer. Over time, the initially more massive star loses mass and becomes the less massive component, while retaining its more evolved internal structure. Consequently, the currently less massive star appears more evolved, located above the TAMS line (\citealt{1971ARAA...9..183P}). This reversed mass ratio is a result of binary interaction rather than isolated stellar evolution. The distinct position of TIC 434222993, where the hotter and less massive component lies above the TAMS line, further suggests the influence of non-conservative mass transfer or angular momentum loss affecting the system's evolutionary path (\citealt{2008MNRAS.390.1577G}).

Figure \ref{4diagrams} also shows the positions of the systems relative to the $T_h-M_m$ theoretical fit from the study by \cite{2024RAA....24e5001P}, where $T_h$ refers to the hotter star in the system and $M_m$ to the more massive component. The positions of these stars are consistent with the $T_h-M_m$ diagram, based on their light curve solutions and estimated absolute parameters. Furthermore, the target systems' position is depicted in the $logM_{tot}-logJ_0$ diagram (Figure \ref{4diagrams}), indicating that they lie within the contact binary systems region.

(D) We compared the results obtained from the BSN application with those generated by the Python implementation of PHOEBE and found them to be very similar (Figure \ref{lc}). To evaluate the degree of similarity between the synthetic light curves produced by the BSN application and the PHOEBE code, we computed the Standard Deviation (STD) of their point-by-point residuals. This statistical measure quantifies how closely the two curves match each other. The STD was calculated using the following equation:

\begin{equation}
\texttt{STD} = \sqrt{ \frac{1}{n} \sum_{i=1}^{n} (r_i - \bar{r})^2 },
\label{eq:std_residuals}
\end{equation}

\noindent where \( r_i = y_i - \hat{y}_i \) represents the residuals between the two synthetic flux values at each time point, and \( \bar{r} \) is the mean of the residuals. A lower value of \( \mathrm{STD}_{\mathrm{res}} \) indicates a higher level of similarity between the two light curves. As a result, the STD values of 0.0069, 0.0065, 0.0065, and 0.0082 were obtained for the systems TIC 406453269, TIC 423750361, TIC 434222993, and TIC 71743300, respectively, for the 100 computational phases. These low STD values demonstrate that the synthetic light curves generated by the two programs exhibit minimal deviations and are in strong numerical agreement. Since the BSN application and PHOEBE adopt different approaches to modeling the geometric structure of binary systems, slight differences in the resulting synthetic light curves are to be expected. PHOEBE employs a simplified spot model based on the \cite{1971ApJ...166..605W} method to define surface elements, whereas the BSN application calculates the three-dimensional shapes of the stars using the method described by \cite{1984ApJS...55..551M}. In addition, both PHOEBE and the BSN application utilize the \texttt{emcee} package for performing MCMC computations. However, due to the BSN application's development and implementation in modern .NET technologies, it achieves fast MCMC (\citealt{2023MNRAS.525.4596D}).

(E) To compare the execution speed of the MCMC-based light curve modeling in the BSN application with that of the Python-based PHOEBE code (version 2.4.9), we ensured that all experimental conditions and parameters were kept identical across both platforms. The tests were conducted on several typical consumer-grade computers, all of which yielded almost consistent results. To provide a quantitative and reproducible comparison, we present one representative test case in detail. In this setup, the number of parameters was identical for both programs, and the algorithms were executed on the same dataset, targeting the same four systems under study. In the MCMC algorithm, it is necessary to perform theoretical light curve modeling a number of times equal to the product of the number of iterations and the number of walkers. Since both PHOEBE and BSN use a similar algorithm for MCMC sampling, the difference in MCMC speed between the two mainly arises from the faster theoretical light curve modeling in the BSN application. A computational phase of 40 was selected for both implementations, and although the role of reflection in contact-based systems is typically minimal, it was included in both programs to ensure consistency. The computer used featured an Intel(R) Core(TM) i7-1065G7 CPU @ 1.30GHz, operating at 1498 MHz, with four physical cores and eight logical processors. The results showed that the BSN application executed the light curve modeling in just 0.2 s, whereas PHOEBE required 8 s for the same task, indicating that BSN was over 40 times faster in this test. Notably, in analyses where reflection effects are excluded, the BSN application achieved an approximately 100-fold increase in the speed of  synthetic light curve generation compared to PHOEBE. These findings underscore the significant performance advantage of the BSN application in executing MCMC under controlled, equivalent conditions.

\vspace{6pt}
\section{Future Works}
- We aim to develop versions of the BSN application compatible with other operating systems.

- We plan to extend the BSN application to support additional photometric systems such as the Sloan Digital Sky Survey (SDSS) and Gaia in future versions. Incorporating these passbands will broaden the applicability of the BSN application, enabling users to perform transformations and analyses across a wider range of astronomical surveys and datasets.

- We will publicly release the second version of this application.

\vspace{6pt}
\section{Acknowledgments}
The BSN application is an effort by the BSN project (\url{https://bsnp.info/}) in the field of contact binary systems. We thank the Raderon AI laboratory who helped at the beginning of this journey. We also thank the members of the Société Astronomique de France (SAF)'s commission des étoiles doubles for their continuous testing of this application.

\vspace{0.6cm}
\bibliography{REFS}{}
\bibliographystyle{aasjournal}

\end{document}